\documentclass[preprint,3p,times]{elsarticle}
\usepackage{amsmath}
\usepackage{graphicx}
\usepackage{epstopdf}
\usepackage{multirow}
\usepackage{amssymb}
\usepackage{footnote}
\usepackage{epsfig}
\usepackage{caption}
\usepackage{url}
\usepackage{rotating}
\usepackage{pdflscape}

\newcommand{\be}{\begin{equation}}
\newcommand{\ee}{\end{equation}}
\def\bea{\begin{eqnarray}} 
\def\eea{\end{eqnarray}} 
\newcommand{\bnel}{\mbox{$\bar{\nu}_e$} }
\newcommand{\natNd}{\mbox{$^{\mathrm{nat}}$Nd}}

\begin{document}
%\begin{frontmatter} 

\title{Excitation functions of proton-induced reactions on natural Nd and production of radionuclides relevant for double beta decay: Completing measurement in 5--35 MeV energy range}

\author[1]{O. Lebeda\corref{cor1}}
\ead{lebeda@ujf.cas.cz}
\author[2]{V. Lozza}
\author[2]{J. Petzoldt\fnref{fn2}}
\author[1]{J. \v{S}tursa}
\author[1]{V. Zdychov\'{a}}
\author[2]{K. Zuber}
\address[1]{Nuclear Physics Institute AS CR, v.v.i., Husinec-\v{R}e\v{z} 130, 250 68 \v{R}e\v{z}, Czech Republic}
\address[2]{Institut f\"{u}r Kern und Teilchenphysik, Technische Universit\"{a}t Dresden, Zellescher Weg 19, 01069 Dresden, Germany}
\cortext[cor1]{Corresponding author: O. Lebeda}
\fntext[fn2]{Present address: OncoRay, National Center for Radiation Research in Oncology, 01307 Dresden, Germany}

\begin{abstract}
Cross-sections for the proton-induced reactions on natural neodymium in energy regions 5--10 MeV and 30--35 MeV were measured using the cyclotron U-120M at the Nuclear Physics Institute at \v{R}e\v{z} near Prague. This measurement completes the investigation previously done in the 10--30 MeV energy range. Results revealed practical production thresholds and secondary maxima and minima in the excitation functions. It allowed for more appropriate calculation of thick target yields and production rates of many longer-lived radionuclides potentially disturbing the search for neutrinoless double beta decay. Measured cross-sections are consistent with our previously published data.
\end{abstract}


%\end{frontmatter}
%\journal{Nuclear Physics A}
\makeatletter
\def\ps@pprintTitle{%
 \let\@oddhead\@empty
 \let\@evenhead\@empty
 \def\@oddfoot{}%
 \let\@evenfoot\@oddfoot}
\makeatother
\maketitle
\section{Introduction}

In the last 80 years, since the neutrino was firstly postulated by Pauli in 1930, the knowledge in neutrino physics has rapidly increased. In addition to its discovery, a non-vanishing rest mass of neutrinos and their mixing could be established. However, the results obtained 
by neutrino oscillations do not allow an absolute neutrino mass measurement.
Furthermore, one of the most interesting open questions is whether the neutrinos are Dirac, or Majorana particles. Both questions can be tackled in neutrinoless double beta decay search \cite{Rod}. It is described as a nuclear decay, where the mass number does not change while the atomic number changes by two units. In addition to the standard weak process resulting in two electrons and two $\bnel$ in the final state, only two electrons should be observed in the neutrinoless decay channel if neutrinos are Majorana particles. This is to be manifested as a peak in the sum energy spectrum of the electrons  at the Q-value of the nuclear transition. \\
$^{150}$Nd is one of the most promising candidates for neutrinoless double beta decay search with its very large Q-value of  3371.38\,$\pm$\,0.20 keV \cite{Kol}  and a natural abundance of 5.6\,\%. It is currently considered to be used by the DCBA(MTD) \cite{Ish12} and SuperNEMO \cite{Arn10} experiments.\\
Despite of the large Q-value exceeding most of the background components due to natural radioactivity, it is important to identify all the radionuclides that can directly or indirectly fall in the energy region where the extremely rare signal is expected. Some of these radionuclides could be produced by cosmogenic neutron and proton activation of neodymium itself. It is, therefore, important to estimate production rate of potential radionuclides formed by cosmic rays in order to define the maximum allowed time for neodymium exposure on Earth's surface, and potentially necessary cooling down times or purification factors before deploying the isotope in the experiment. Cosmogenic proton activation of \natNd~on  Earth's surface results, among others, in $^{143}$Pm, $^{144}$Pm, $^{146}$Pm, $^{148}$Pm, $^{148m}$Pm, $^{147}$Nd and $^{139}$Ce. These isotopes are of concern because of their long half-life, or their high Q-value, or feeding of high Q-value isotopes.\\ The production cross-sections for proton activation in the energy region 10--30 MeV have been carefully investigated in \cite{Leb12}\cite{Leb12E}. However, it is desirable to complete the measurement in the low (5--10 MeV) and high (30--35 MeV) energy region to study the shape of excitation functions down to their thresholds and up to maximum available proton energy at the cyclotron U-120M. Moreover, the missing cross-sections allow to specify more accurately production rates of the majority of the longer-lived activation products.\\
In Table \ref{tab::reaction_channels} we summarize reaction channels that are missing in Table II of \cite{Leb12} or are open in the higher energy region.

\begin{table}[Ht]
\captionsetup{labelsep=newline, justification=raggedright, singlelinecheck=false}
\caption{Activation channels of the proton-induced reactions on natural neodymium not shown in \cite{Leb12}. Values are adopted from \cite{Pri14}.}\label{tab::reaction_channels}
\begin{center}
\begin{tabular}{llc}
\hline
Radionuclide	&		Reaction channels		&		Q-value (MeV)		\\
\hline									
$^{140}$Pm + $^{140m}$Pm  &  $^{143}$Nd(p,4n)		&	$	-30.790	$	\\
$^{141}$Pm 	&	$^{144}$Nd(p,4n)		&	$	-28.221	$	\\
	&		$^{145}$Nd(p,5n)	 &	$	-33.977	$	\\
$^{144}$Pm	&		$^{148}$Nd(p,5n)		&	$	-29.060	$	\\
$^{146}$Pm	&		$^{150}$Nd(p,5n)		&	$	-27.293  $	\\
$^{140}$Nd	&		$^{145}$Nd(p,t3n)		&	$	-29.053	$	\\
$^{141}$Nd + $^{141m}$Nd	 &	$^{146}$Nd(p,t3n)		&	$	-28.607	$	\\
$^{138m}$Pr	&	$^{145}$Nd(p,$\alpha$4n)		&	$	-25.721	$	\\
$^{140}$Pr	&		$^{146}$Nd(p,$\alpha$3n)		&	$	-15.584	$	\\
	&		$^{148}$Nd(p,$\alpha$5n)		&	$	-28.209	$	\\
$^{142}$Pr + $^{142m}$Pr	&		$^{148}$Nd(p,$\alpha$3n)		&	$	-12.969	$	\\
	&		$^{150}$Nd(p,$\alpha$5n)		&	$	-25.383	$	\\
$^{139}$Ce	&	$^{142}$Nd(p,$^{3}$Hep)		&	$	-13.931	$	\\
	&		$^{143}$Nd(p,$\alpha$p)		&	$	+0.523	$	\\
	&		$^{144}$Nd(p,$\alpha$d)		&	$	-5.069	$	\\
	&		$^{145}$Nd(p,$\alpha$t)		&	$	-4.567	$	\\
	&		$^{146}$Nd(p,$\alpha$tn)		&	$	-12.133	$	\\
\hline 
\end{tabular}
\end{center}
\end{table}

\section{Experimental setup}

The excitation functions were measured by the usual activation stacked-foil technique on the external beam of the isochronous cyclotron U-120M in the Nuclear Physics Institute in \v{R}e\v{z} as described in \cite{Leb12}. For the high energy region (30--35 MeV), four natural neodymium targets of 10$\times $10 mm$^{2}$ area were prepared from commercially available Nd foils (99.9\% purity, AlfaAesar). The foil thickness was measured using low energy gamma ray ($^{241}$Am source) absorption both in Dresden and in \v{R}e\v{z} with two different set-ups. Due to the large distance between the Am source and the Nd foil, the attenuation coefficient with coherent scattering has been used for the measurement in Dresden, while in \v{R}e\v{z}, due to the shorter source-foil distance, the attenuation coefficient has been used without coherent scattering \cite{xcom}. The two results have shown very good agreement with an averaged value of 104.21\,$\pm$\,0.51\,$\mu$m. For the low energy range (5--10\,MeV), thinner Nd foils have been purchased from EspiMetals (99.9\% purity) in order to achieve acceptably low proton beam energy decrease in a foil. Four thin targets of 10$\times$10 mm$^{2}$ area were obtained from a 25$\times$25 mm$^{2}$ foil. The nominal thickness of the foil was 25\,$\mu$m, however, the measurement with the $^{241}$Am source has revealed target thicknesses to be about 20 \% lower and also slightly differing from foil to foil. It has been decided to determine the thickness of each foil separately by precise measurement of their dimensions and by measuring the mass of neodymium in order to reduce its uncertainty. The mass was determined via neutron activation analysis (NAA) using neodymium trioxide (99.99 \%, REactonT, UK) calcinated shortly before use at 900\,$^{o}$C for 1 hour as external standard. The NAA was performed on the thin neodymium targets several months after the proton irradiation. Its results are shown in Table \ref{tab::thickness}. \\
In order to prevent the oxidation of the neodymium foils, both the thin and the thick targets were covered with a Parylene-C layer of 2--3\,$\mu$m thickness.\\
The targets were fixed in a water-cooled holder designed for irradiating stacks of foils. Each stack consisted of four neodymium foils of the same thickness interleaved with titanium beam monitors and copper degraders of various thickness. The actual thicknesses of the Ti (AlfaAesar) and Cu (GoodFellow) foils were determined repeatedly by precise weighing of discs or rectangular foil's pieces of known dimensions. In both cases, the thickness of commercially available foils was very uniform with estimated uncertainty $<$ 1\%. The foils were ordered as follows:
\begin{enumerate}
\item  Titanium (thickness 12.11\,$\mu$m) acting as a beam monitor;
\item Parylene-coated neodymium foil;
\item Copper beam energy degrader (21.2 or 109.0\,$\mu$m);
\item  Titanium (thickness 12.11\,$\mu$m) acting as a beam monitor;
\item Parylene-coated neodymium foil;
\item Copper beam energy degrader (21.2 or 109.0\,$\mu$m);
\item  Titanium (thickness 12.11\,$\mu$m) acting as a beam monitor;
\item Parylene-coated neodymium foil;
\item Copper beam energy degrader (21.2 or 54.5\,$\mu$m);
\item Titanium (thickness 12.11\,$\mu$m) acting as a beam monitor;
\item Parylene-coated neodymium foil;
\item Thick silver foil acting as a beam stop, directly cooled by water.
\end{enumerate}

\begin{table}[tbp]
\captionsetup{labelsep=newline, justification=raggedright, singlelinecheck=false}
\begin{center}
\caption{Thickness of the thin Nd foils from EspiMetals used in the stack for 5--10\,MeV region as measured by the NAA.}\label{tab::thickness}
\begin{tabular}{cc}
\hline
Foil no.	&		Thickness ($\mu$m)	\\
\hline									
1         &        19.17	$\pm$	0.42	\\
2	&	19.62	$\pm$	0.40	\\
3	&	18.86	$\pm$	0.44	\\
4	&	18.92	$\pm$	0.44	\\
\hline 
\end{tabular}
\end{center}
\end{table}

The low energy stack was irradiated for 2 hours with protons of entrance energy equal to 9.50\,$\pm$\,0.20\,MeV and beam current equal to 0.410 $\mu$A,
while the high energy stack only for 1 hour with protons of 35.79\,$\pm$\,0.20\,MeV entrance energy and of 0.370 $\mu$A beam current. The beam energy was calculated from precisely measured beam orbit position (uncertainty ca 0.20\,MeV) and its decrease in the stack using the code SRIM \cite{cycl07}\cite{srim}.  The beam current was calculated from the activity of $^{48}$V produced in the Ti monitors via $^{nat}$Ti(p,x)$^{48}$V reaction and recommended cross-sections for this reaction, cf.  \cite{Leb12}.

\section{Data processing}

Activity of a single radionuclide in a given foil at the end of bombardment was obtained as described in \cite{Leb12}. Only single well resolved $\gamma$-peak of high intensity was used for activity calculation (cf. Table III and paragraphs devoted to particular radionuclides in \cite{Leb12})---it makes further correction of the published cross-sections due to future intensity value up-date very easy. Correction was applied when a radionuclide is directly produced in nuclear reactions and simultaneously formed by decay of another radionuclide born in the target. This is particularly the case of $^{141}$Nd (formed by decay of $^{141}$Pm) and $^{149}$Pm (formed by decay of $^{149}$Nd).\\
If the activity of a parent nuclide cannot be directly measured (it is e.g. too short-lived or it has no sufficiently intense gamma lines), only cumulative cross-sections can be obtained. In one case ($^{140}$Nd, T$_{1/2}$\,=\,3.37\,d), activity of the long-lived parent with missing gamma emission was deduced from the activity of its short-lived daughter ($^{140}$Pr, T$_{1/2}$\,=\,3.4\,min) that reaches very soon transient equilibrium with the parent.
Decay chains occurring in the Nd activation are provided in \cite{Leb12}.\\ 
The overall cross-section uncertainty was calculated as a square root of sum of squares of the following sources of uncertainty: \\
\begin{itemize}
\item beam current measurement: 10 \%. This value is conservative since recommended cross-sections for monitoring reactions are provided without uncertainties \cite{Tar01};
\item detection efficiency: 3\,\%;
\item thick foil thickness: 0.5\,\%;
\item thin foil thickness: 2.4\,\%;
\item gamma line intensity: $\leq$ 10\,\% (mostly below 3\,\%);
\item net peak area: 0.4--22.2\,\% (mostly below 2\,\%). If the net peak area is corrected for contribution of indirectly born radionuclide, the overall absolute peak uncertainty is calculated as the square root of the sums of squares of the absolute uncertainties of the subtracted quantities (e.g. $^{141}$Nd and $^{149}$Pm);
\item overall uncertainty: 11--25\,\% (mostly between 11 and 12.5\,\%)
\end{itemize}

\section{Results and Discussion}

\begin{figure}[]	
		\begin{center}
		\includegraphics[angle=0,width=130mm]{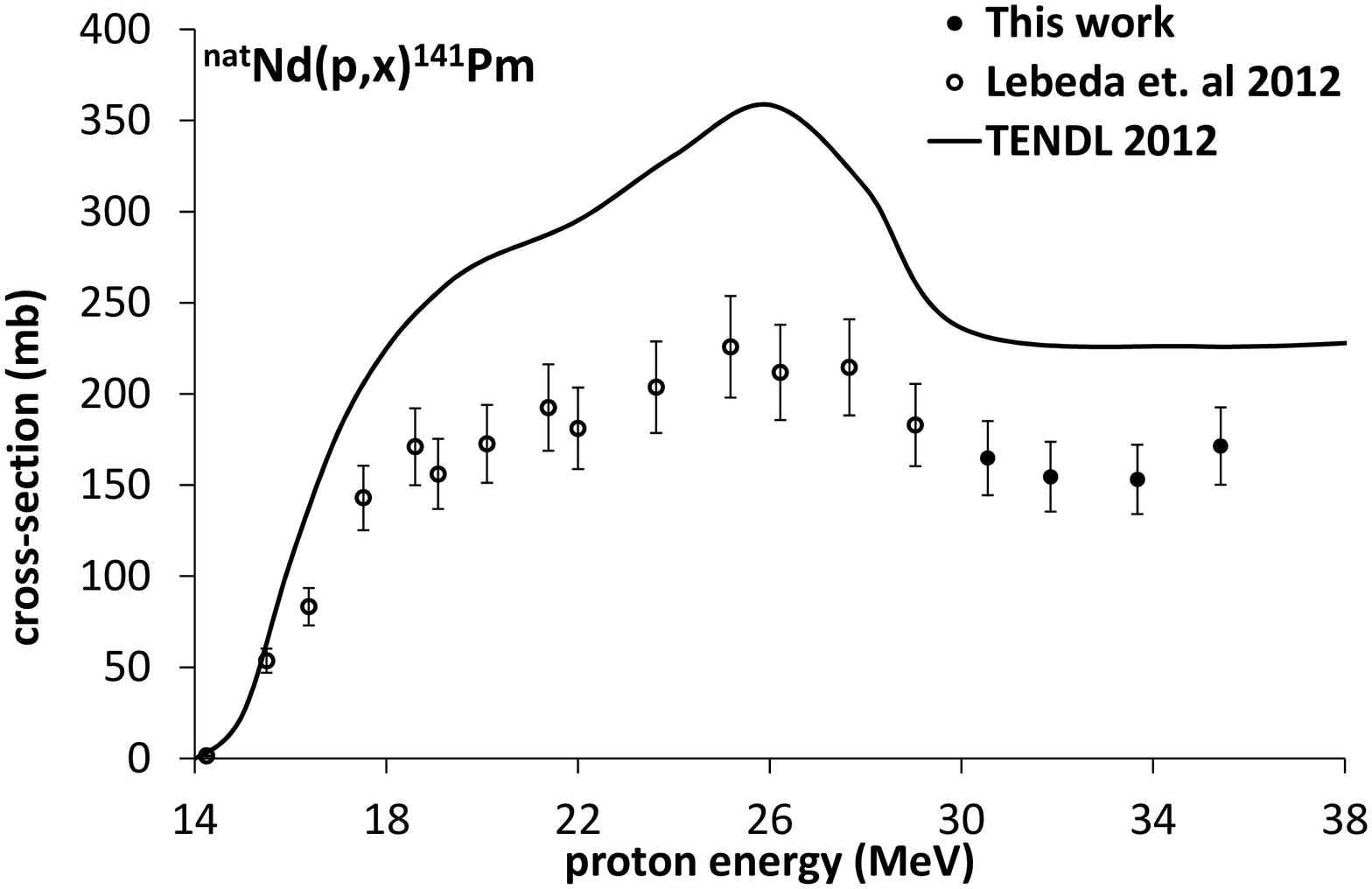}
		\caption{Experimental cross-sections for the \natNd(p,x)$^{141}$Pm reactions compared with data from \cite{Leb12} (open circles) and the TENDL2012 library (solid line).}
		\label{fig::xs1}
		\end{center}
\end{figure}

\begin{figure}[]	
		\begin{center}
		\includegraphics[angle=0,width=130mm]{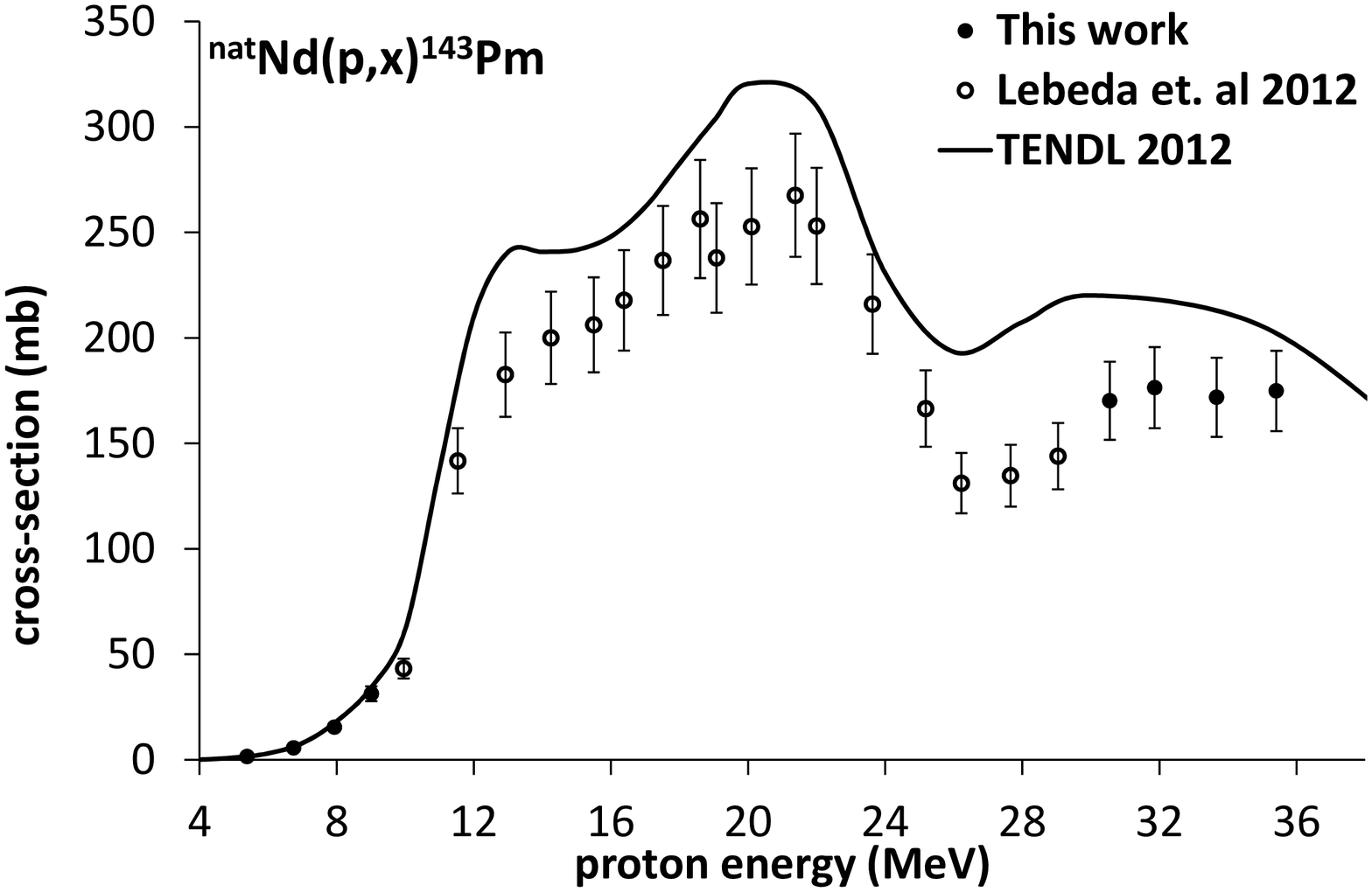}
		\caption{Experimental cross-sections for the \natNd(p,x)$^{143}$Pm reactions compared with data from \cite{Leb12} (open circles) and the TENDL2012 library (solid line).}
		\label{fig::xs2}
		\end{center}
\end{figure}

\begin{figure}[]	
		\begin{center}
		\includegraphics[angle=0,width=130mm]{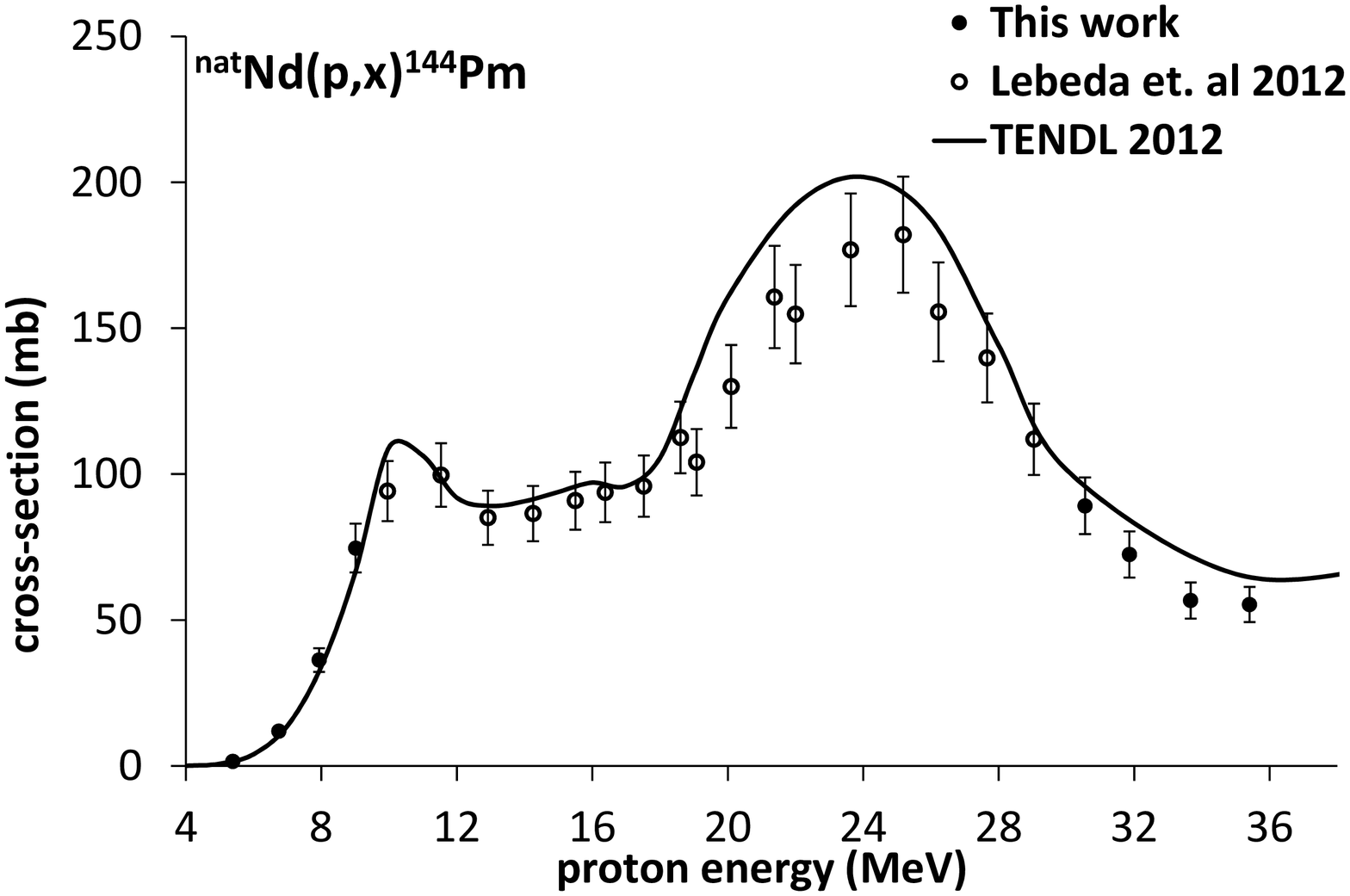}
		\caption{Experimental cross-sections for the \natNd(p,x)$^{144}$Pm reactions compared with data from \cite{Leb12} (open circles) and the TENDL2012 library (solid line).}
		\label{fig::xs3}
		\end{center}
\end{figure}

\begin{figure}[]	
		\begin{center}
		\includegraphics[angle=0,width=130mm]{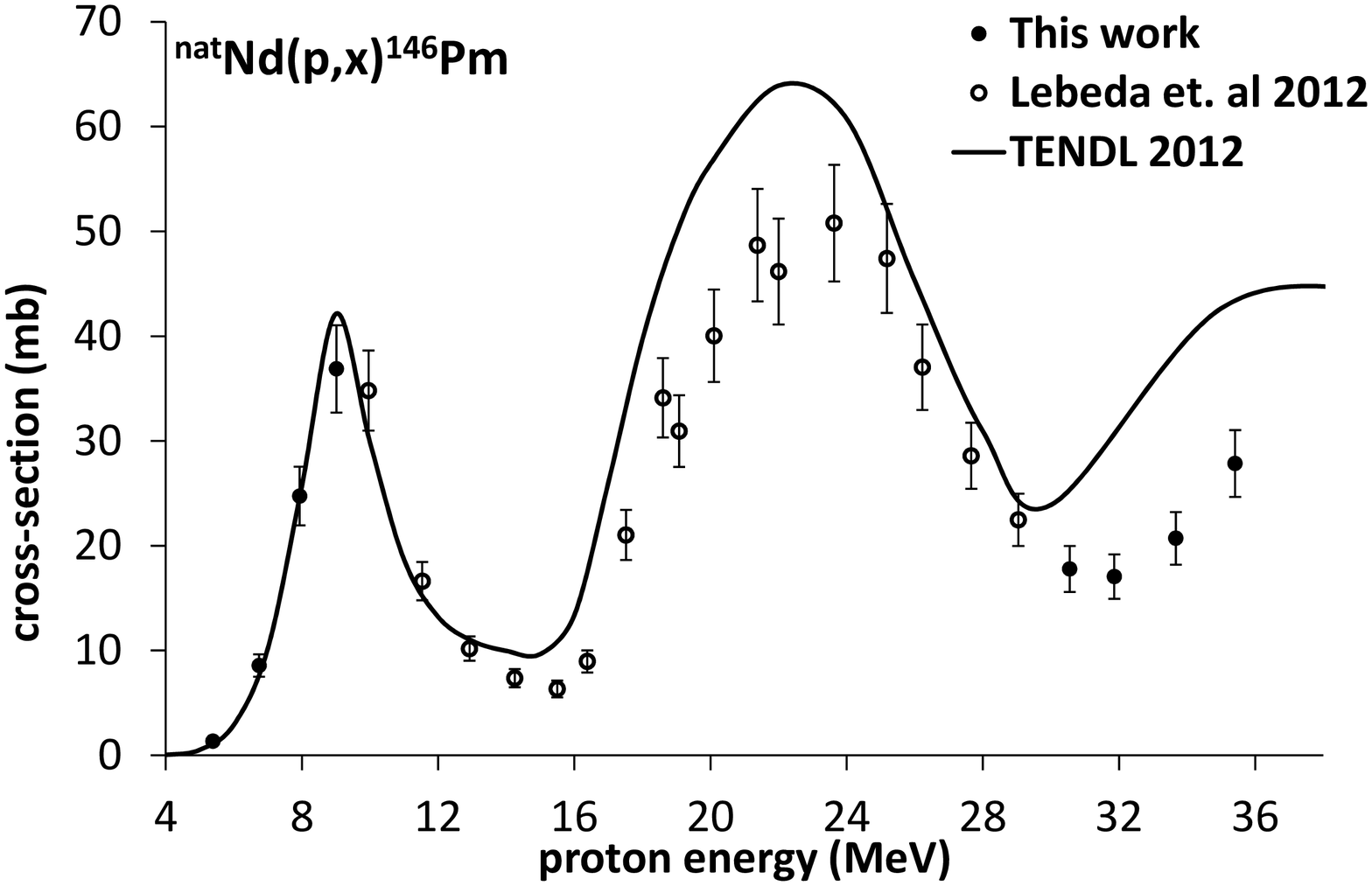}
		\caption{Experimental cross-sections for the \natNd$(p,x)^{146}$Pm reactions compared with data from \cite{Leb12} (open circles) and the TENDL2012 library (solid line).}
		\label{fig::xs4}
		\end{center}
\end{figure}

\begin{figure}[]	
		\begin{center}
		\includegraphics[angle=0,width=130mm]{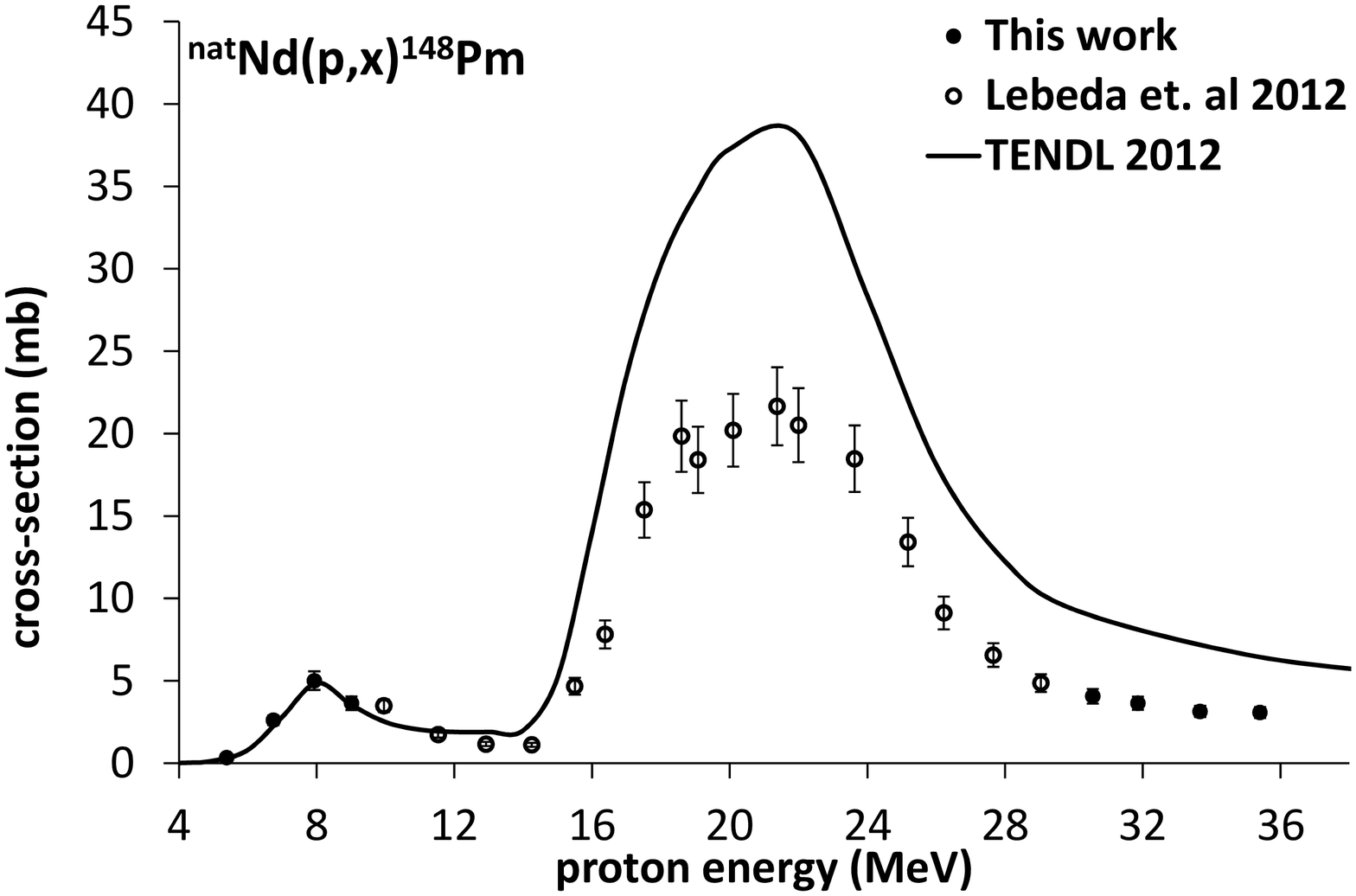}
		\caption{Experimental cross-sections for the \natNd(p,x)$^{148}$Pm reactions compared with data from \cite{Leb12} (open circles) and the TENDL2012 library (solid line).}
		\label{fig::xs5}
		\end{center}
\end{figure}

\begin{figure}[]	
		\begin{center}
		\includegraphics[angle=0,width=130mm]{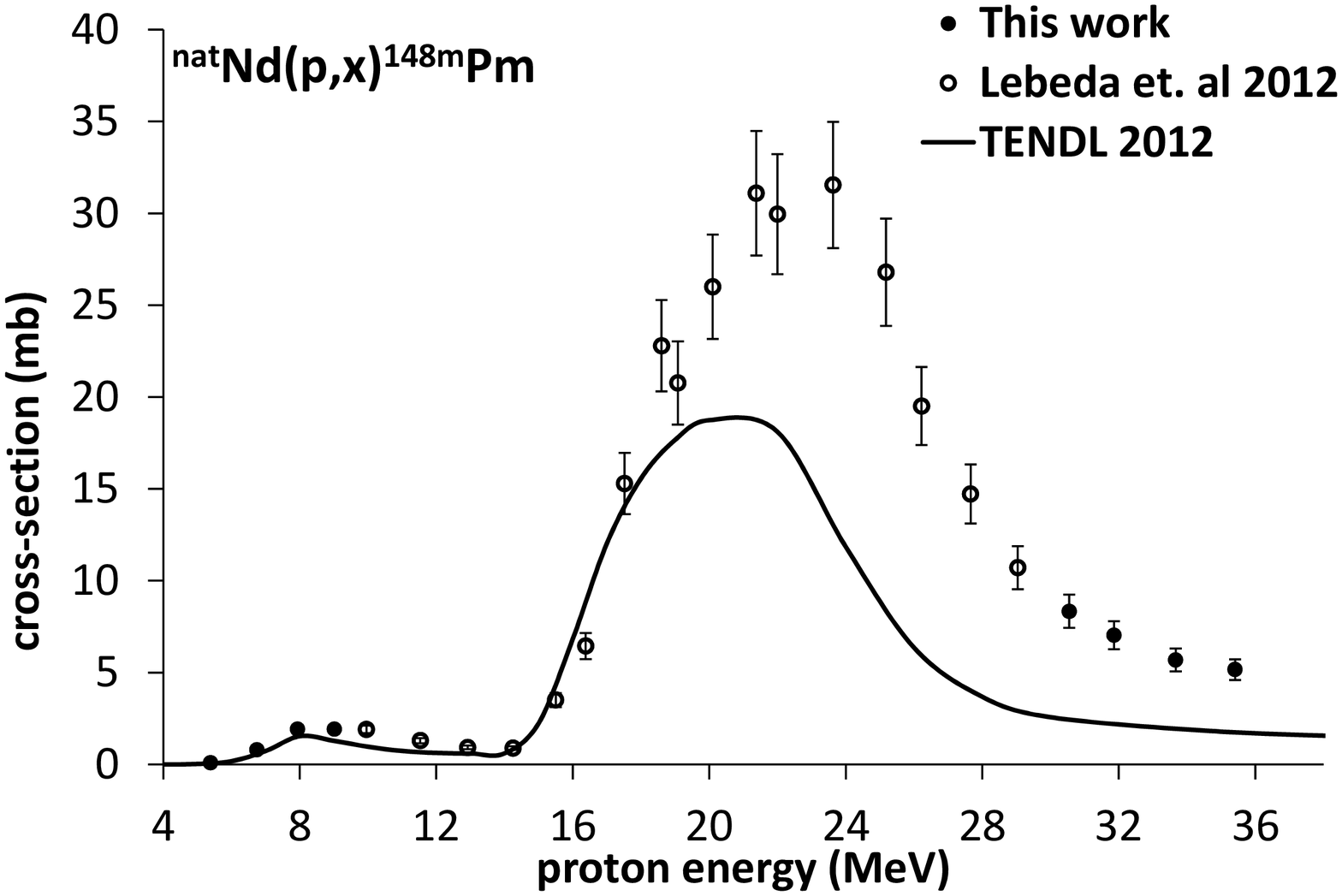}
		\caption{Experimental cross-sections for the \natNd(p,x)$^{148m}$Pm reactions compared with data from \cite{Leb12} (open circles) and the TENDL2012 library (solid line).}
		\label{fig::xs6}
		\end{center}
\end{figure}

\begin{figure}[]	
		\begin{center}
		\includegraphics[angle=0,width=130mm]{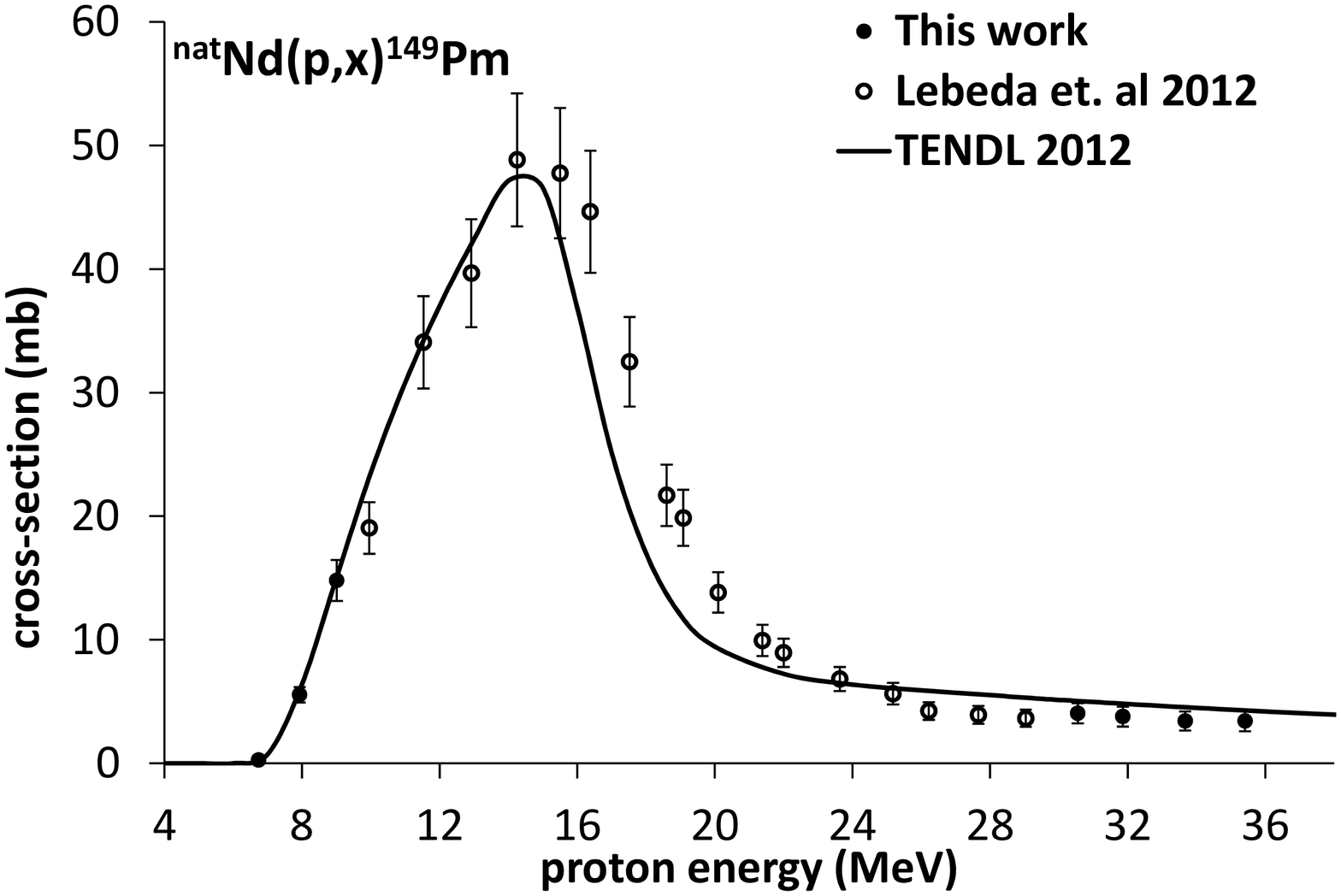}
		\caption{Experimental cross-sections for the \natNd(p,x)$^{149}$Pm reactions compared with data from \cite{Leb12}\cite{Leb12E} (open circles) and the TENDL2012 library (solid line).}
		\label{fig::xs7}
		\end{center}
\end{figure}

\begin{figure}[]	
		\begin{center}
		\includegraphics[angle=0,width=130mm]{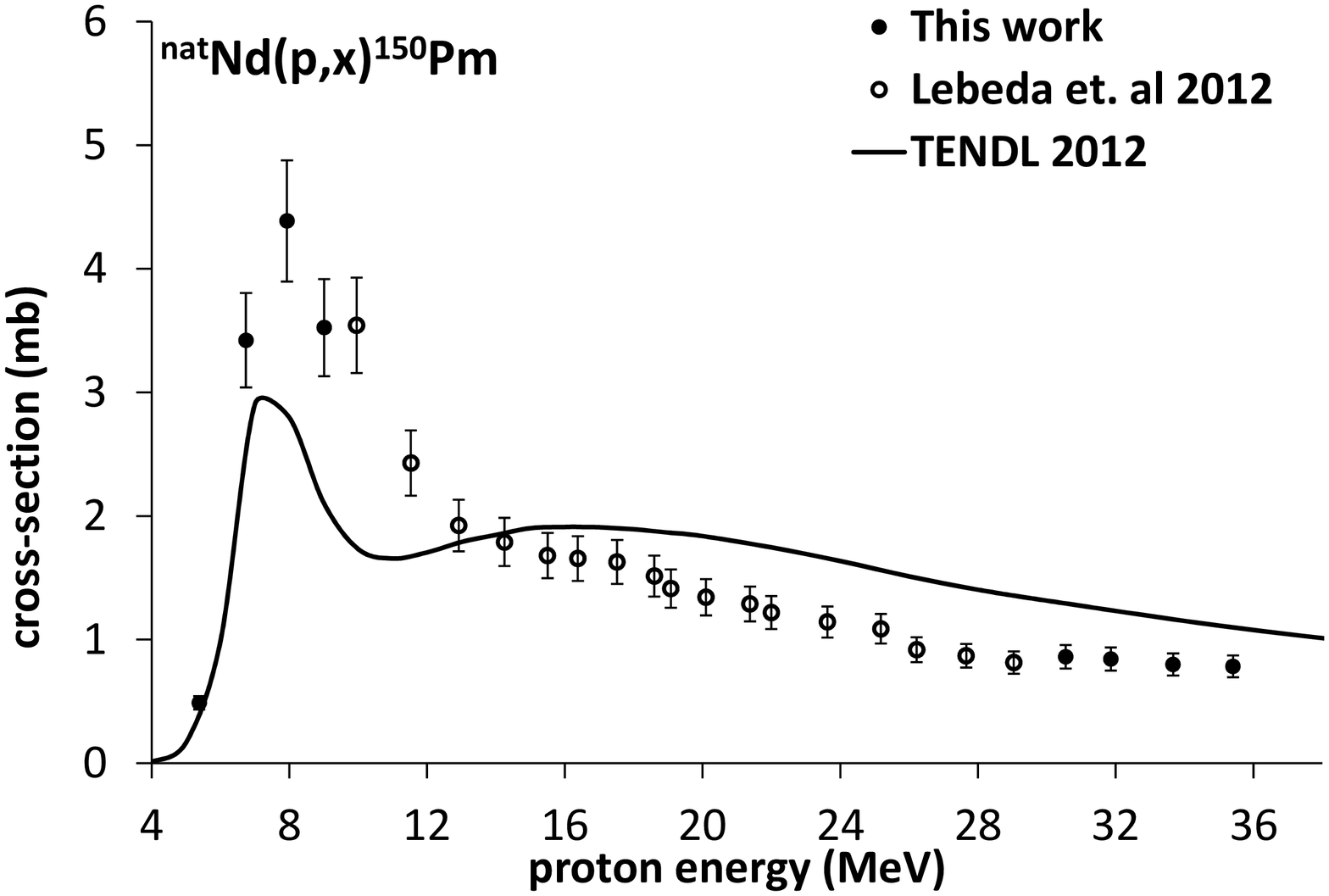}
		\caption{Experimental cross-sections for the \natNd(p,x)$^{150}$Pm reaction compared with data from \cite{Leb12} (open circles) and the TENDL2012 library (solid line).}
		\label{fig::xs8}
		\end{center}
\end{figure}

\begin{figure}[]	
		\begin{center}
		\includegraphics[angle=0,width=130mm]{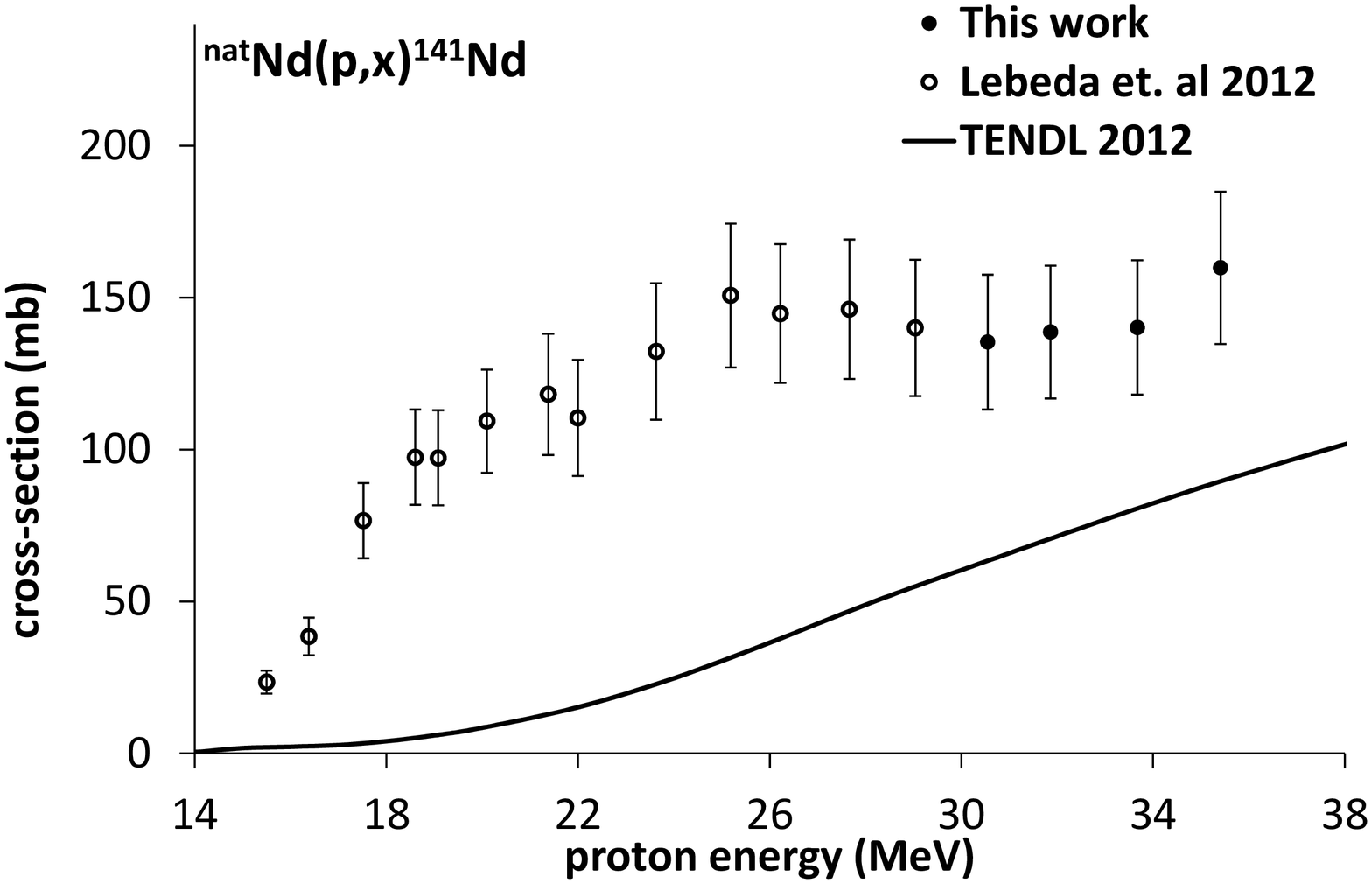}
		\caption{Experimental cross-sections for the \natNd(p,x)$^{141}$Nd reactions compared with data from \cite{Leb12}\cite{Leb12E} (open circles) and the TENDL2012 library (solid line).}
		\label{fig::xs9}
		\end{center}
\end{figure}

\begin{figure}[]	
		\begin{center}
		\includegraphics[angle=0,width=130mm]{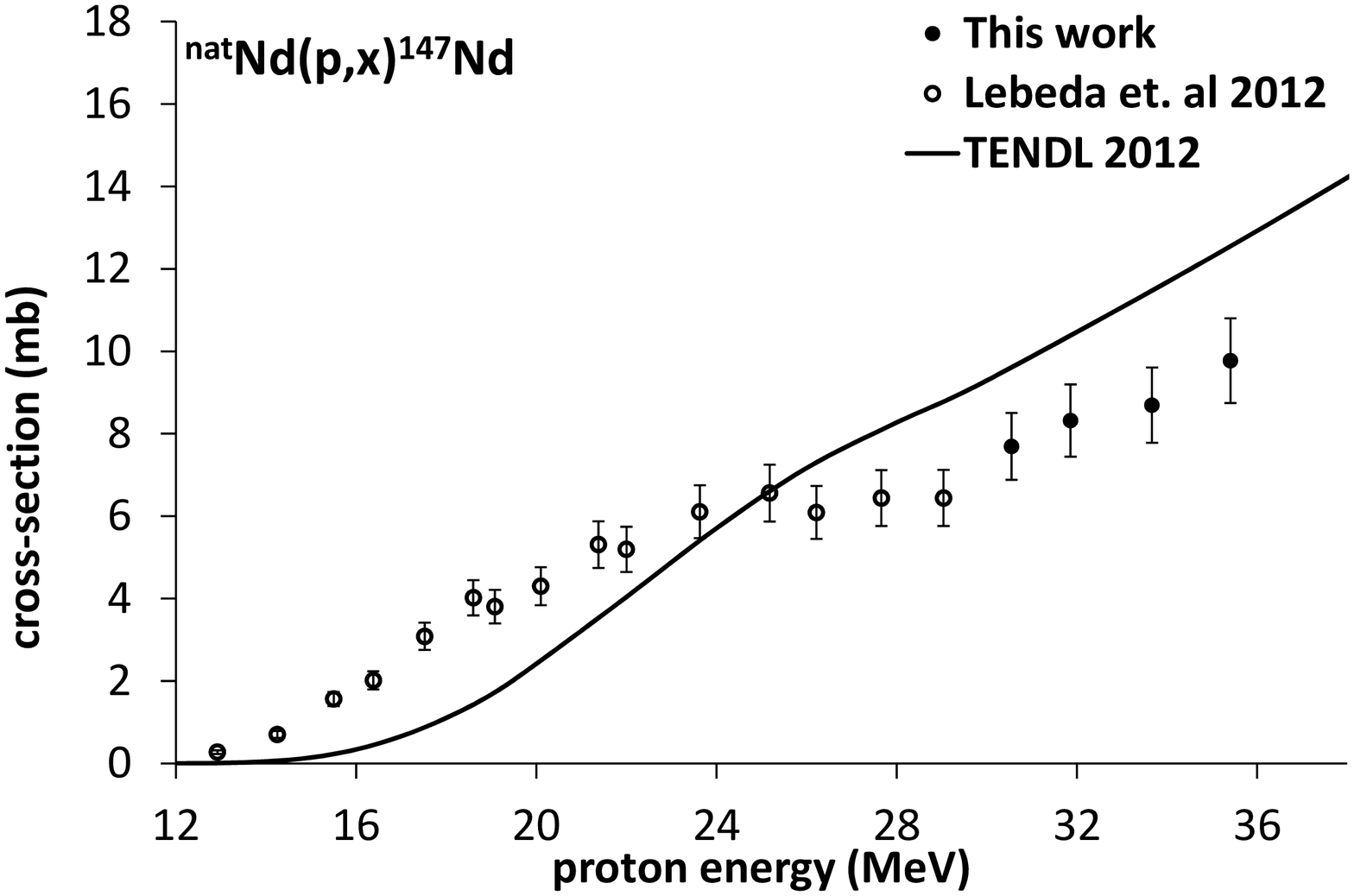}
		\caption{Experimental cross-sections for the \natNd(p,x)$^{147}$Nd reactions compared with data from \cite{Leb12} (open circles) and the TENDL2012 library (solid line).}
		\label{fig::xs10}
		\end{center}
\end{figure}

\begin{figure}[]	
		\begin{center}
		\includegraphics[angle=0,width=130mm]{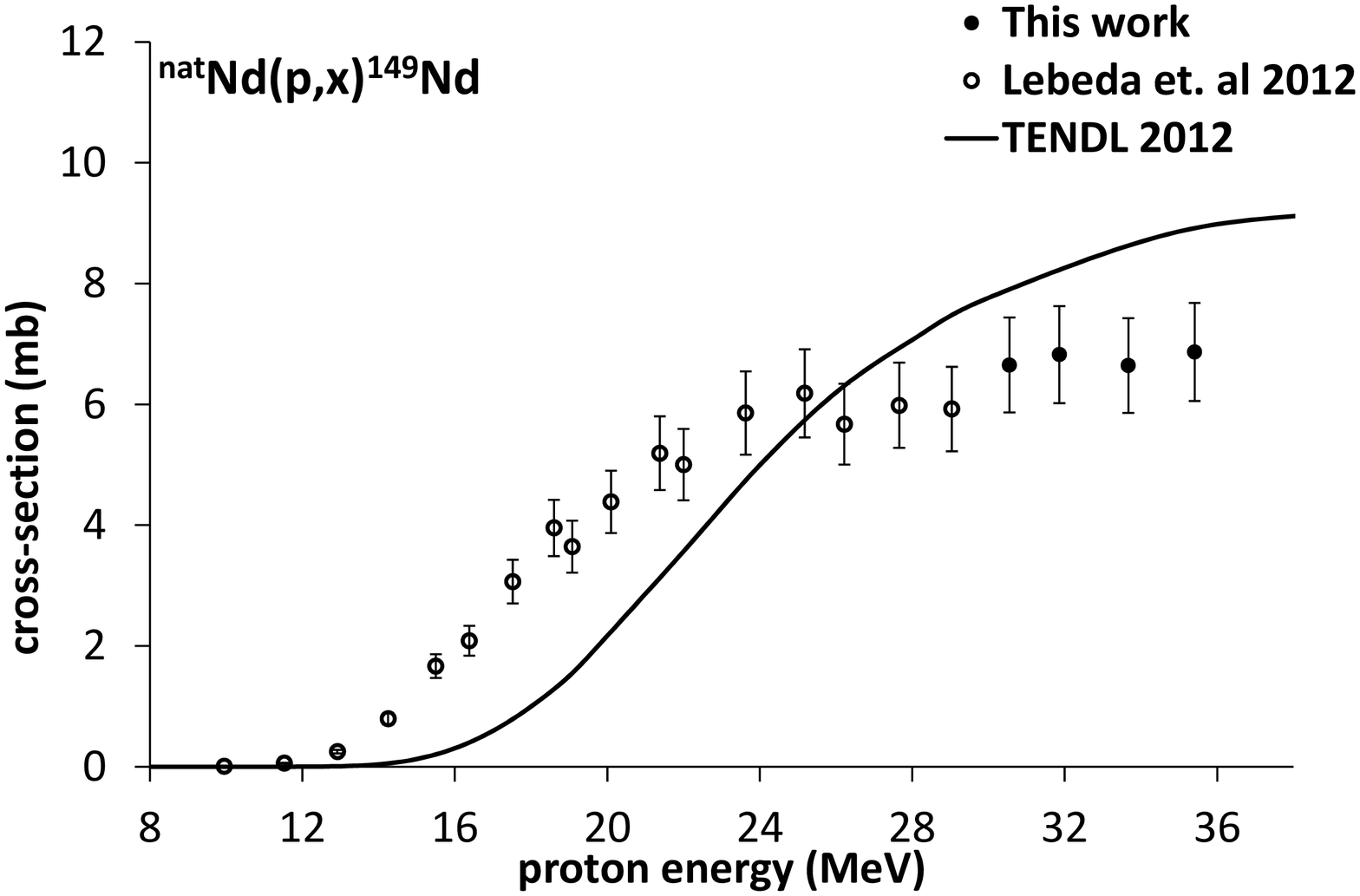}
		\caption{Experimental cross-sections for the \natNd(p,x)$^{149}$Nd reactions compared with data from \cite{Leb12} (open circles) and the TENDL2012 library (solid line).}
		\label{fig::xs11}
		\end{center}
\end{figure}

\begin{figure}[]	
		\begin{center}
		\includegraphics[angle=0,width=130mm]{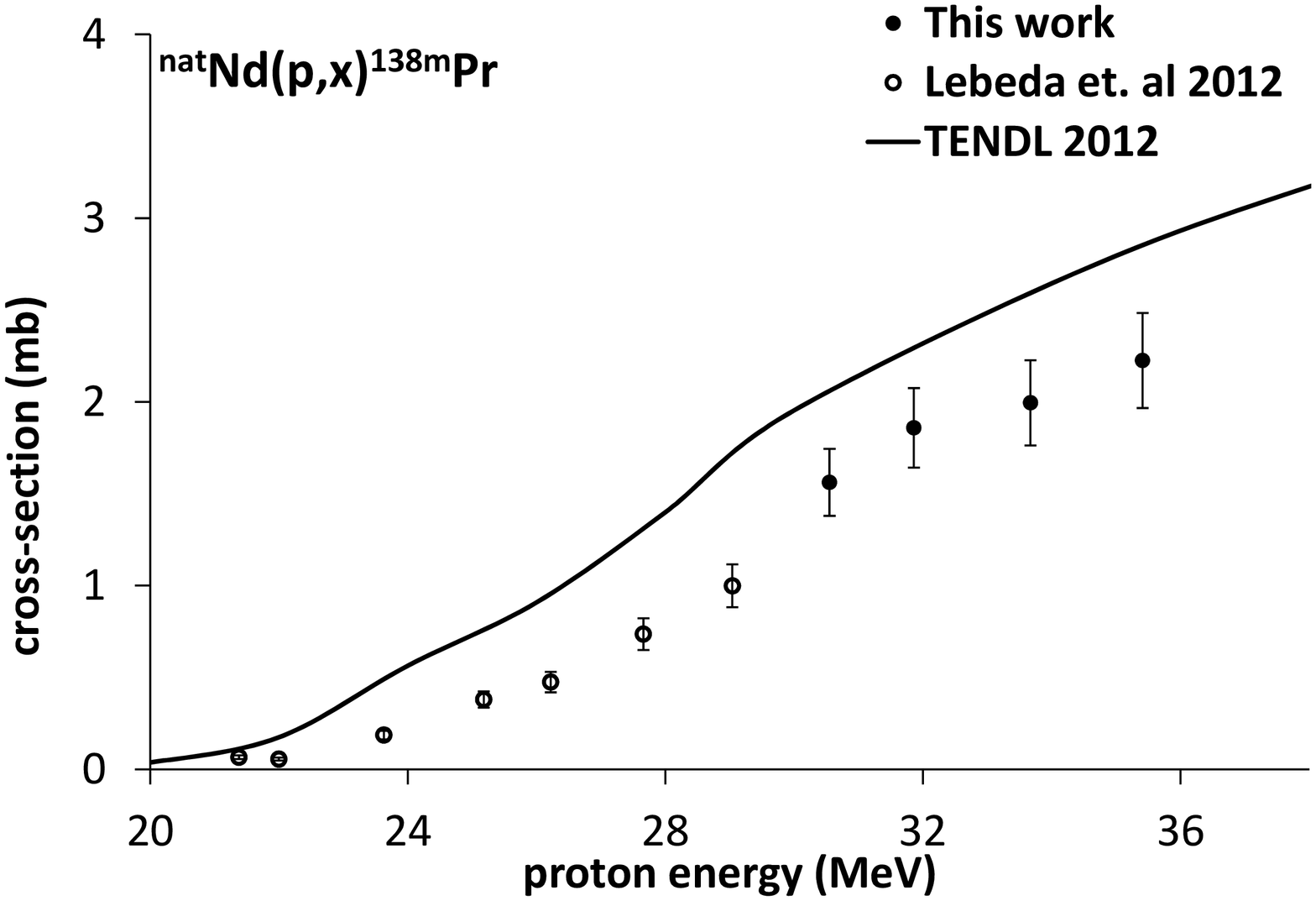}
		\caption{Experimental cross-sections for the \natNd(p,x)$^{138m}$Pr reactions compared with data from \cite{Leb12} (open circles) and the TENDL2012 library (solid line).}
		\label{fig::xs12}
		\end{center}
\end{figure}

\begin{figure}[]	
		\begin{center}
		\includegraphics[angle=0,width=130mm]{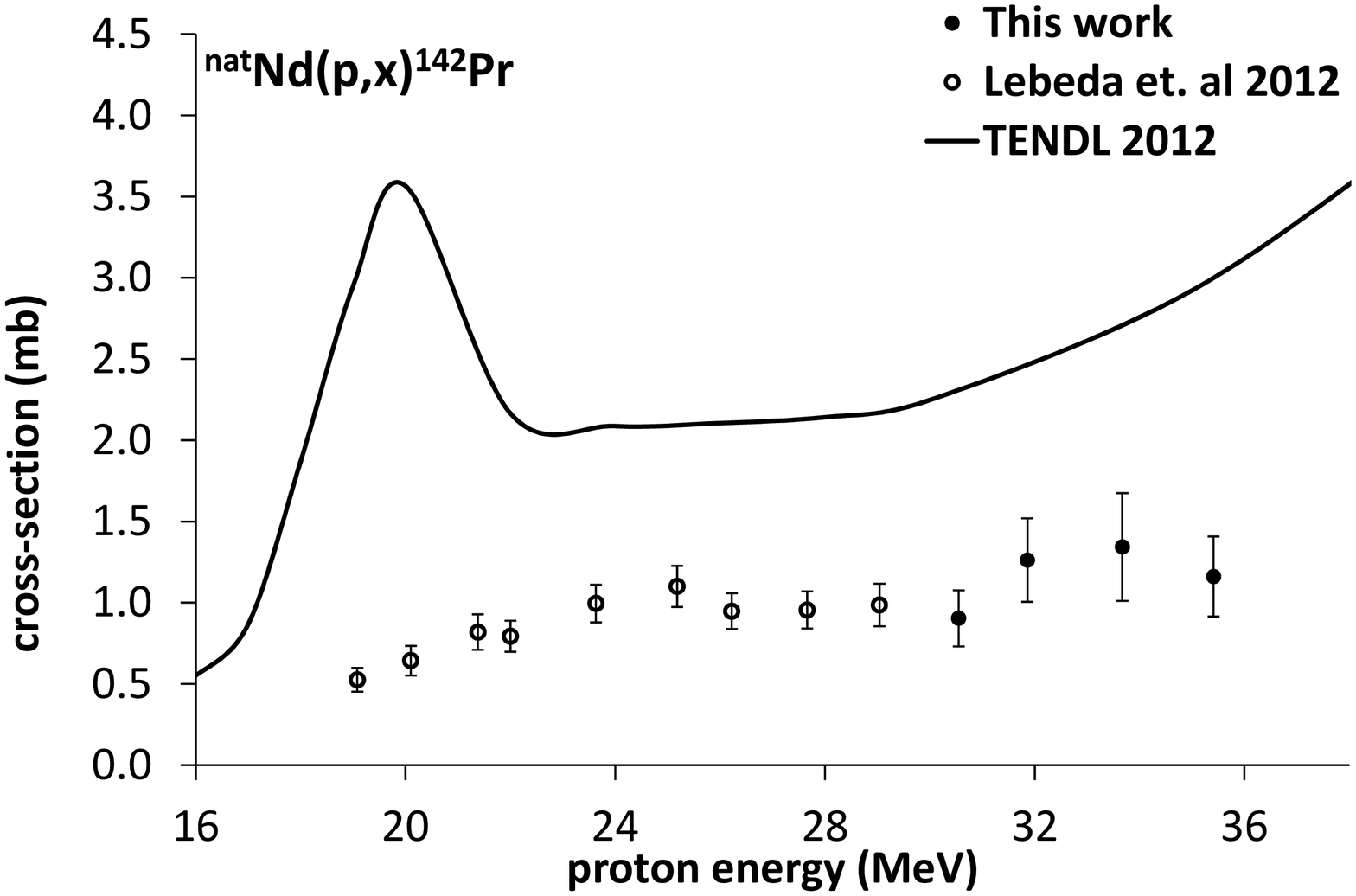}
		\caption{Experimental cross-sections for the \natNd(p,x)$^{142}$Pr reactions compared with data from \cite{Leb12} (open circles) and the TENDL2012 library (solid line).}
		\label{fig::xs14}
		\end{center}
\end{figure}

Measured cross-sections are displayed in Tables \ref{tab::xs_1} and \ref{tab::xs_2}. The values obtained in \cite{Leb12}\cite{Leb12E} are also shown for comparison. Most of the measured data are elemental cross-sections, in some cases both elemental and cumulative or only cumulative cross-sections are provided (cumulative data are denoted by right upper index {\it cum}). For $^{150}$Pm and $^{149}$Pm, elemental cross-sections can be converted easily to isotopic cross-sections, since they are produced solely on $^{150}$Nd. Selected cross-sections are plotted in Figs. \ref{fig::xs1}--\ref{fig::xs14} along with the calculated excitation functions based on the TENDL2012 \cite{talys} library. 

\section{Discussion}
\begin{figure}[]	
		\begin{center}
		\includegraphics[angle=0,width=130mm]{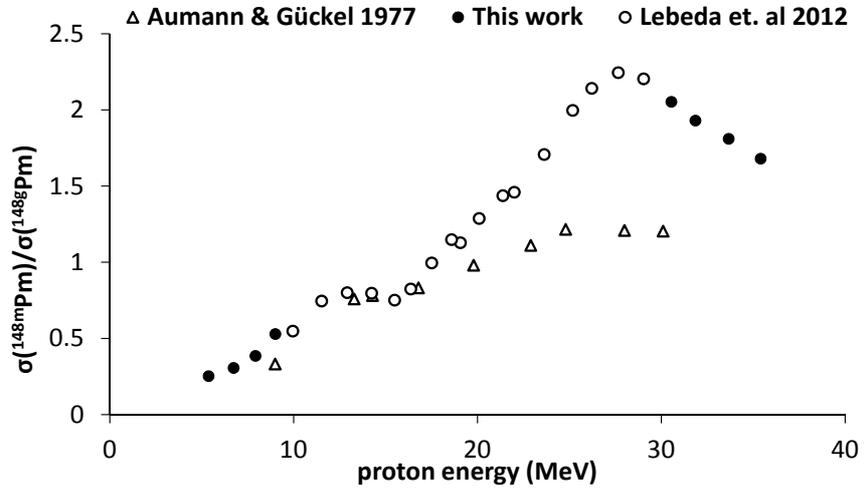}
		\caption{Experimental isomeric cross-section ratios for the $^{148}$Nd(p,n) reaction measured by Aumann and G\"uckel (1977) and for the \natNd(p,x) reactions measured in \cite{Leb12} (open circles) and this work (full circles).}
		\label{fig::Pm148GSvsM}
		\end{center}
\end{figure}

\begin{figure}[ht]	
		\begin{center}
		\includegraphics[angle=0,width=130mm]{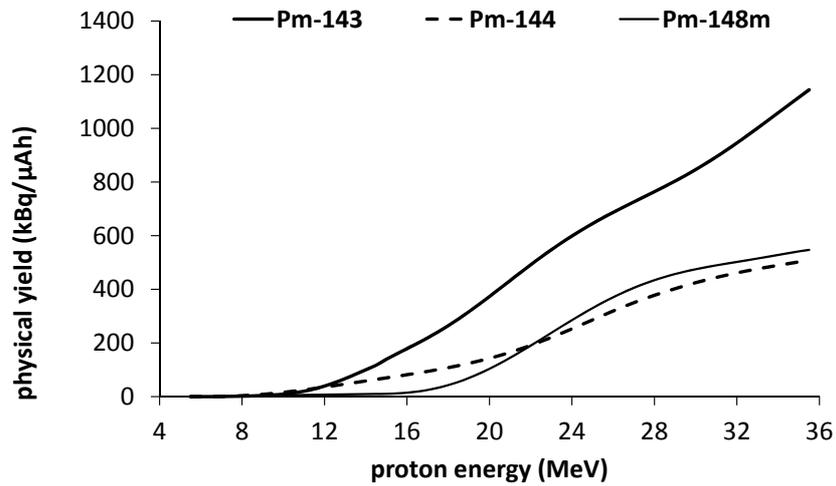}
		\caption{Physical thick target yield of $^{143}$Pm, $^{144}$Pm, and $^{148m}$Pm in the \natNd(p,x) reactions for $E_{\text{out}} = 5.5\,$MeV.}
		\label{fig::TY1}
		\end{center}
\end{figure}

\begin{figure}[Ht]	
		\begin{center}
		\includegraphics[angle=0,width=130mm]{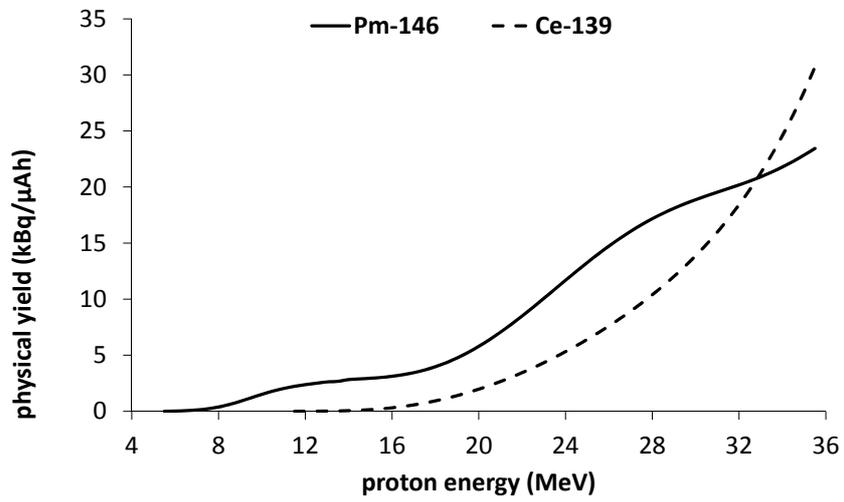}
		\caption{Physical thick target yield of $^{146}$Pm, and $^{139}$Ce in the \natNd(p,x) reactions for $E_{\text{out}} = 5.5\,$MeV.}
		\label{fig::TY2}
		\end{center}
\end{figure}

The measurement of the cross-sections for formation of radionuclides by proton-induced reactions on natural neodymium was performed in the energy range of 5--10 MeV and 30--35 MeV in order to complete previously published data. New cross-sections revealed production thresholds, in particular for the majority of  long-lived radionuclides, and extended experimental excitation functions for 5 MeV towards higher energies. It allowed for more appropriate calculation of the thick target yields from practical reaction thresholds onwards.\\
Measured cross-sections were compared to the results published in \cite{Leb12}\cite{Leb12E} and to the TENDL2012 library \cite{talys}. New data are in very good agreement with the previously measured values, while some discrepancies between them and the prediction adopted from TENDL2012 persist. As it was already noted in \cite{Leb12}, for the isomeric pair of $^{148m}$Pm and $^{148}$Pm, the measured ratio of the metastable and ground states is about 2, while the predicted value is below 1. The comparison with data of Aumann and G\"{u}ckel \cite{Aum77} is shown in Fig. \ref{fig::Pm148GSvsM}. The difference above 17 MeV seems to be due to the contribution of the $^{150}$Nd(p,3n) reaction, since data of \cite{Aum77} were measured only for the $^{148}$Nd(p,n) reaction with use of enriched $^{148}$Nd target.\\
From a low background experiment point of view, the most critical radionuclides are those having long half-life or a high enough Q-value to interfere with the measurement. It concerns in particular $^{143}$Pm, $^{144}$Pm, $^{146}$Pm, $^{148m}$Pm and $^{139}$Ce. Physical thick target yields for these five radionuclides based on our extended  experimental cross-sections are shown in Fig. \ref{fig::TY1} and \ref{fig::TY2}. The used threshold was 5.5 MeV. \\
An estimate of the expected production rates at sea level for the long-lived radionuclides in the energy range of 5--35 MeV is displayed in Table \ref{tab::prodrate}. The proton flux was adopted from \cite{Bar06}. Production rates are given in decays/kg/d of a neodymium target of natural isotopic composition. For comparison, values based on predicted cross-sections adopted from TENDL2012 library are also displayed. The ratio is in a good agreement with our data  published previously in \cite{Leb12}\cite{Leb12E}. 

\section*{Acknowledgements}
We thank the crew of the cyclotron U-120M for performing irradiations and energy calculations from the orbit position, Alex Fauler for the preparation of the parylene target coating and Prof. Jan Ku\v{c}era for neutron activation analysis of the decayed thin Nd targets. The work was supported by the Deutsche Forschungsgemeinschaft (DFG) ZU 123/5, by the Academy of Sciences of the Czech Republic under the NPI research plan AV0Z10480505 and by the CANAM project funded by the Ministry of Education, Youth and Sports of the Czech Republic (project No. LM2011019).

\section*{References}
\bibliography{bibliography_R2.bib}
\bibliographystyle{elsarticle-num}

\begin{landscape}
\begin{table}[t]
\captionsetup{labelsep=newline, justification=raggedright, singlelinecheck=false}
\centering
\caption{Measured cross-sections for the formation of $^{141}$Pm, $^{143}$Pm, $^{144}$Pm, $^{146}$Pm, $^{148}$Pm, $^{148m}$Pm, $^{149}$Pm, $^{149}$Pm$^{cum}$, $^{150}$Pm. $^{\dagger}$ = this work, $^{\ddagger}$ = \cite{Leb12}\cite{Leb12E}.}\label{tab::xs_1}
%\begin{center}
\begin{tabular}{cccccccccc}
\hline
E$_{p}$ (MeV)	&	\multicolumn{9}{c}{Cross section (mb)} \\ \cline{2-10}
	&	$^{141}$Pm			&	$^{143}$Pm			&	$^{144}$Pm			&	$^{146}$Pm			&	$^{148}$Pm			&	$^{148m}$Pm			&	$^{149}$Pm			&	$^{149}$Pm$^{cum}$			&	$^{150}$Pm			\\ \hline
35.41 $\pm$0.20$^{\dagger}$	&	171	$\pm$	21	&	175	$\pm$	19	&	55.3	$\pm$	6.1	&	27.8	$\pm$	3.2	&	3.07	$\pm$	0.34	&	5.16	$\pm$	0.56	&	3.43	$\pm$	0.83	&	10.5	$\pm$	1.22	&	0.784	$\pm$	0.088	\\
33.67 $\pm$0.21$^{\dagger}$	&	153	$\pm$	19	&	172	$\pm$	19	&	56.7	$\pm$	6.2	&	20.7	$\pm$	2.5	&	3.14	$\pm$	0.35	&	5.69	$\pm$	0.62	&	3.42	$\pm$	0.77	&	10.3	$\pm$	1.16	&	0.799	$\pm$	0.089	\\
31.86 $\pm$0.21$^{\dagger}$	&	155	$\pm$	19	&	176	$\pm$	19	&	72.5	$\pm$	7.9	&	17.1	$\pm$	2.1	&	3.64	$\pm$	0.40	&	7.03	$\pm$	0.77	&	3.78	$\pm$	0.80	&	10.8	$\pm$	1.22	&	0.842	$\pm$	0.093	\\
30.55 $\pm$0.22$^{\dagger}$	&	165	$\pm$	20	&	170	$\pm$	19	&	89.1	$\pm$	9.7	&	17.8	$\pm$	2.2	&	4.06	$\pm$	0.44	&	8.34	$\pm$	0.91	&	4.04	$\pm$	0.82	&	10.9	$\pm$	1.23	&	0.860	$\pm$	0.095	\\
29.04 $\pm$0.26$^{\ddagger}$	&	183	$\pm$	23	&	144	$\pm$	16	&	112	$\pm$	12	&	22.5	$\pm$	2.5	&	4.86	$\pm$	0.53	&	10.7	$\pm$	1.2	&	3.64	$\pm$	0.69	&	9.76	$\pm$	1.08	&	0.814	$\pm$	0.090	\\
27.66 $\pm$0.26$^{\ddagger}$	&	215	$\pm$	26	&	135	$\pm$	15	&	140	$\pm$	15	&	28.6	$\pm$	3.1	&	6.56	$\pm$	0.72	&	14.7	$\pm$	1.6	&	3.91	$\pm$	0.72	&	10.1	$\pm$	1.1	&	0.869	$\pm$	0.096	\\
26.22 $\pm$0.27$^{\ddagger}$	&	212	$\pm$	26	&	131	$\pm$	14	&	156	$\pm$	17	&	37.0	$\pm$	4.1	&	9.11	$\pm$	1.00	&	19.5	$\pm$	2.1	&	4.23	$\pm$	0.72	&	10.1	$\pm$	1.1	&	0.919	$\pm$	0.101	\\
25.18 $\pm$0.25$^{\ddagger}$	&	226	$\pm$	28	&	167	$\pm$	18	&	182	$\pm$	20	&	47.4	$\pm$	5.2	&	13.4	$\pm$	1.5	&	26.8	$\pm$	2.9	&	5.64	$\pm$	0.88	&	12.0	$\pm$	1.3	&	1.09	$\pm$	0.12	\\
23.63 $\pm$0.27$^{\ddagger}$	&	204	$\pm$	25	&	216	$\pm$	24	&	177	$\pm$	19	&	50.8	$\pm$	5.6	&	18.5	$\pm$	2.0	&	31.5	$\pm$	3.4	&	6.82	$\pm$	0.97	&	12.9	$\pm$	1.4	&	1.14	$\pm$	0.13	\\
22.00 $\pm$0.28$^{\ddagger}$	&	181	$\pm$	22	&	253	$\pm$	28	&	155	$\pm$	17	&	46.2	$\pm$	5.1	&	20.5	$\pm$	2.2	&	30.0	$\pm$	3.3	&	8.93	$\pm$	1.14	&	14.1	$\pm$	1.5	&	1.22	$\pm$	0.13	\\
21.38 $\pm$0.26$^{\ddagger}$	&	193	$\pm$	24	&	268	$\pm$	29	&	161	$\pm$	18	&	48.7	$\pm$	5.4	&	21.7	$\pm$	2.4	&	31.1	$\pm$	3.4	&	9.94	$\pm$	1.26	&	15.3	$\pm$	1.7	&	1.29	$\pm$	0.14	\\
20.10 $\pm$0.27$^{\ddagger}$	&	173	$\pm$	21	&	253	$\pm$	28	&	130	$\pm$	14	&	40.0	$\pm$	4.4	&	20.2	$\pm$	2.2	&	26.0	$\pm$	2.8	&	13.8	$\pm$	1.64	&	18.4	$\pm$	2.0	&	1.34	$\pm$	0.15	\\
19.08 $\pm$0.28$^{\ddagger}$	&	156	$\pm$	19	&	238	$\pm$	26	&	104	$\pm$	11	&	30.9	$\pm$	3.4	&	18.4	$\pm$	2.0	&	20.8	$\pm$	2.3	&	19.9	$\pm$	2.3	&	23.6	$\pm$	2.6	&	1.41	$\pm$	0.15	\\
18.60 $\pm$0.26$^{\ddagger}$	&	171	$\pm$	21	&	256	$\pm$	28	&	113	$\pm$	12	&	34.1	$\pm$	3.8	&	19.8	$\pm$	2.2	&	22.8	$\pm$	2.5	&	21.7	$\pm$	2.5	&	25.8	$\pm$	2.8	&	1.51	$\pm$	0.17	\\
17.52 $\pm$0.27$^{\ddagger}$	&	143	$\pm$	18	&	237	$\pm$	26	&	95.9	$\pm$	10.5	&	21.0	$\pm$	2.4	&	15.4	$\pm$	1.7	&	15.3	$\pm$	1.7	&	32.5	$\pm$	3.6	&	35.7	$\pm$	3.9	&	1.63	$\pm$	0.18	\\
16.38 $\pm$0.28$^{\ddagger}$	&	83.2	$\pm$	10.3	&	218	$\pm$	24	&	93.7	$\pm$	10.2	&	8.95	$\pm$	1.05	&	7.82	$\pm$	0.86	&	6.44	$\pm$	0.70	&	44.6	$\pm$	4.9	&	46.8	$\pm$	5.1	&	1.66	$\pm$	0.18	\\
15.50 $\pm$0.26$^{\ddagger}$	&	53.6	$\pm$	6.7	&	206	$\pm$	23	&	90.9	$\pm$	9.9	&	6.32	$\pm$	0.79	&	4.67	$\pm$	0.51	&	3.51	$\pm$	0.38	&	47.8	$\pm$	5.3	&	49.5	$\pm$	5.4	&	1.68	$\pm$	0.18	\\
14.25 $\pm$0.28$^{\ddagger}$	&	1.43	$\pm$	0.21	&	200	$\pm$	22	&	86.5	$\pm$	9.4	&	7.36	$\pm$	0.86	&	1.11	$\pm$	0.12	&	0.886	$\pm$	0.098	&	48.8	$\pm$	5.4	&	49.7	$\pm$	5.5	&	1.79	$\pm$	0.19	\\
12.92 $\pm$0.29$^{\ddagger}$	&				&	183	$\pm$	20	&	85.1	$\pm$	9.3	&	10.2	$\pm$	1.2	&	1.16	$\pm$	0.13	&	0.928	$\pm$	0.103	&	39.7	$\pm$	4.4	&	39.9	$\pm$	4.4	&	1.92	$\pm$	0.21	\\
11.53 $\pm$0.27$^{\ddagger}$	&				&	142	$\pm$	15	&	99.7	$\pm$	11	&	16.6	$\pm$	1.8	&	1.73	$\pm$	0.19	&	1.29	$\pm$	0.14	&	34.1	$\pm$	3.7	&	34.1	$\pm$	3.7	&	2.43	$\pm$	0.26	\\
9.96 $\pm$0.30$^{\ddagger}$	&				&	43.2	$\pm$	4.7	&	94.2	$\pm$	10.3	&	34.8	$\pm$	3.8	&	3.48	$\pm$	0.38	&	1.91	$\pm$	0.21	&	19.0	$\pm$	2.1	&	19.0	$\pm$	2.1	&	3.54	$\pm$	0.39	\\
9.01 $\pm$0.21$^{\dagger}$	&				&	31.3	$\pm$	3.5	&	74.6	$\pm$	8.4	&	36.9	$\pm$	4.2	&	3.64	$\pm$	0.41	&	1.92	$\pm$	0.22	&	14.8	$\pm$	1.7	&	14.8	$\pm$	1.7	&	3.52	$\pm$	0.39	\\
7.93 $\pm$0.23$^{\dagger}$	&				&	15.5	$\pm$	1.7	&	36.3	$\pm$	4.1	&	24.7	$\pm$	2.8	&	5.01	$\pm$	0.56	&	1.93	$\pm$	0.22	&	5.54	$\pm$	0.63	&	5.54	$\pm$	0.63	&	4.39	$\pm$	0.49	\\
6.74 $\pm$0.26$^{\dagger}$	&				&	5.58	$\pm$	0.64	&	11.9	$\pm$	1.3	&	8.57	$\pm$	1.1	&	2.60	$\pm$	0.29	&	0.797	$\pm$	0.089	&	0.261	$\pm$	0.033	&	0.261	$\pm$	0.033	&	3.42	$\pm$	0.38	\\
5.38 $\pm$0.30$^{\dagger}$	&				&	1.58	$\pm$	0.19	&	1.52	$\pm$	0.17	&	1.35	$\pm$	0.27	&	0.345	$\pm$	0.039	&	0.0871	$\pm$	0.011	&				&				&	0.489	$\pm$	0.055	\\\hline 
\end{tabular}
%\end{center}
\end{table}
\end{landscape}

\begin{landscape}
\begin{table}[t]
\captionsetup{labelsep=newline, justification=raggedright, singlelinecheck=false}
\centering
\caption{Measured cross-sections for the formation of $^{140}$Nd$^{cum}$, $^{141}$Nd, $^{147}$Nd, $^{149}$Nd, $^{138m}$Pr, $^{139}$Pr, $^{142}$Pr, $^{139}$Ce$^{cum}$. $^{\dagger}$ = this work, $^{\ddagger}$ = \cite{Leb12}\cite{Leb12E}.}\label{tab::xs_2}
\begin{tabular}{cccccccc}
\hline 
E$_{p}$ (MeV)	&	\multicolumn{7}{c}{Cross section (mb)} \\ \cline{2-8}
	&	$^{140}$Nd$^{cum}$			&	$^{141}$Nd			&	$^{147}$Nd			&	$^{149}$Nd			&	$^{138m}$Pr			&	$^{142}$Pr			&	$^{139}$Ce$^{cum}$			\\\hline
35.41 $\pm$0.20$^{\dagger}$	&	225	$\pm$	24.7	&	160	$\pm$	25	&	9.77	$\pm$	1.03	&	6.87	$\pm$	0.81	&	2.22	$\pm$	0.26	&	1.16	$\pm$	0.247	&	7.28	$\pm$	1.05	\\
33.67 $\pm$0.21$^{\dagger}$	&	202	$\pm$	22.1	&	140	$\pm$	22	&	8.69	$\pm$	0.91	&	6.64	$\pm$	0.78	&	1.99	$\pm$	0.23	&	1.34	$\pm$	0.331	&	5.22	$\pm$	0.76	\\
31.86 $\pm$0.21$^{\dagger}$	&	175	$\pm$	19.2	&	139	$\pm$	22	&	8.32	$\pm$	0.88	&	6.82	$\pm$	0.80	&	1.86	$\pm$	0.22	&	1.26	$\pm$	0.257	&	4.48	$\pm$	0.65	\\
30.55 $\pm$0.22$^{\dagger}$	&	140	$\pm$	15.4	&	135	$\pm$	22	&	7.69	$\pm$	0.81	&	6.65	$\pm$	0.78	&	1.56	$\pm$	0.18	&	0.904	$\pm$	0.173	&	4.05	$\pm$	0.59	\\
29.04 $\pm$0.26$^{\ddagger}$	&	75.5	$\pm$	8.3	&	140	$\pm$	22	&	6.44	$\pm$	0.68	&	5.92	$\pm$	0.70	&	1.00	$\pm$	0.12	&	0.986	$\pm$	0.131	&	3.08	$\pm$	0.45	\\
27.66 $\pm$0.26$^{\ddagger}$	&	29.0	$\pm$	3.2	&	146	$\pm$	23	&	6.44	$\pm$	0.68	&	5.98	$\pm$	0.71	&	0.736	$\pm$	0.087	&	0.955	$\pm$	0.114	&	2.71	$\pm$	0.39	\\
26.22 $\pm$0.27$^{\ddagger}$	&	4.67	$\pm$	0.56	&	145	$\pm$	23	&	6.09	$\pm$	0.64	&	5.67	$\pm$	0.67	&	0.475	$\pm$	0.056	&	0.948	$\pm$	0.109	&	2.28	$\pm$	0.33	\\
25.18 $\pm$0.25$^{\ddagger}$	&	1.94	$\pm$	0.34	&	151	$\pm$	24	&	6.56	$\pm$	0.69	&	6.18	$\pm$	0.73	&	0.380	$\pm$	0.045	&	1.10	$\pm$	0.127	&	2.34	$\pm$	0.34	\\
23.63 $\pm$0.27$^{\ddagger}$	&	1.20	$\pm$	0.23	&	132	$\pm$	22	&	6.11	$\pm$	0.64	&	5.86	$\pm$	0.69	&	0.186	$\pm$	0.024	&	0.995	$\pm$	0.117	&	2.15	$\pm$	0.31	\\
22.00 $\pm$0.28$^{\ddagger}$	&				&	110.4	$\pm$	19	&	5.19	$\pm$	0.55	&	5.00	$\pm$	0.59	&	0.0564	$\pm$	0.0082	&	0.794	$\pm$	0.096	&	1.86	$\pm$	0.27	\\
21.38 $\pm$0.26$^{\ddagger}$	&				&	118.2	$\pm$	20	&	5.31	$\pm$	0.56	&	5.19	$\pm$	0.61	&	0.0663	$\pm$	0.0097	&	0.819	$\pm$	0.109	&	1.88	$\pm$	0.27	\\
20.10 $\pm$0.27$^{\ddagger}$	&				&	109.3	$\pm$	17	&	4.30	$\pm$	0.46	&	4.39	$\pm$	0.52	&				&	0.643	$\pm$	0.092	&	1.53	$\pm$	0.22	\\
19.08 $\pm$0.28$^{\ddagger}$	&				&	97.3	$\pm$	16	&	3.81	$\pm$	0.41	&	3.64	$\pm$	0.43	&				&	0.525	$\pm$	0.072	&	1.17	$\pm$	0.17	\\
18.60 $\pm$0.26$^{\ddagger}$	&				&	97.5	$\pm$	16	&	4.02	$\pm$	0.43	&	3.95	$\pm$	0.47	&				&				&	1.28	$\pm$	0.19	\\
17.52 $\pm$0.27$^{\ddagger}$	&				&	76.6	$\pm$	12.4	&	3.08	$\pm$	0.33	&	3.06	$\pm$	0.36	&				&				&	0.942	$\pm$	0.139	\\
16.38 $\pm$0.28$^{\ddagger}$	&				&	38.5	$\pm$	6.2	&	2.02	$\pm$	0.22	&	2.09	$\pm$	0.25	&				&				&	0.564	$\pm$	0.085	\\
15.50 $\pm$0.26$^{\ddagger}$	&				&	23.5	$\pm$	3.8	&	1.56	$\pm$	0.17	&	1.66	$\pm$	0.20	&				&				&	0.475	$\pm$	0.071	\\
14.25 $\pm$0.28$^{\ddagger}$	&				&				&	0.703	$\pm$	0.082	&	0.794	$\pm$	0.094	&				&				&	0.232	$\pm$	0.037	\\
12.92 $\pm$0.29$^{\ddagger}$	&				&				&	0.272	$\pm$	0.045	&	0.250	$\pm$	0.030	&				&				&	0.103	$\pm$	0.019	\\
11.53 $\pm$0.27$^{\ddagger}$	&				&				&				&	0.0584	$\pm$	0.0076	&				&				&	0.0475	$\pm$	0.0133	\\
9.96 $\pm$0.30$^{\ddagger}$	&				&				&				&	0.00493	$\pm$	0.0014	&				&				&				\\
9.01 $\pm$0.21$^{\dagger}$	&				&				&				&				&				&				&				\\
7.93 $\pm$0.23$^{\dagger}$	&				&				&				&				&				&				&				\\
6.74 $\pm$0.26$^{\dagger}$	&				&				&				&				&				&				&				\\
5.38 $\pm$0.30$^{\dagger}$	&				&				&				&				&				&				&				\\
 \hline
\end{tabular}
\end{table}
\end{landscape}

\begin{table*}[Htbp]
\captionsetup{labelsep=newline, justification=raggedright, singlelinecheck=false}
\centering
\caption{Calculated production rate of long-lived radionuclides at sea level. The proton flux value was adopted from \cite{Bar06}. The production rate based on the TENDL2012 library has been calculated in the energy range from 5 MeV to 36 MeV.}\label{tab::prodrate}
\begin{tabular}{ccccc}
\hline 							
Radionuclide	&	T$_{1/2}$	&	Production rate exp.	&	Production rate TENDL	&	Ratio	\\
	&	(d)	&	(kg$^{-1}$d$^{-1}$)	&	(kg$^{-1}$d$^{-1}$)	&	TENDL/Exp	\\ \hline
$^{143}$Pm	&	265	&	0.1753	&	0.2187	&	1.25	\\
$^{144}$Pm	&	363	&	0.1092	&	0.1196	&	1.10	\\
$^{146}$Pm	&	2019.83	&	0.0276	&	0.0356	&	1.29	\\
$^{148}$Pm	&	5.37	&	0.0089	&	0.0151	&	1.70	\\
$^{148m}$Pm	&	41.29	&	0.0132	&	0.0064	&	0.48	\\
$^{147}$Nd	&	10.98	&	0.0051	&	0.0052	&	1.02	\\
$^{139}$Ce	&	137.64	&	0.0022	&	0.0028	&	1.26	\\
 \hline 
\end{tabular}
\end{table*}

\end{document}